# Detailed physical property characterization of FeTe$_{1-x}$Se$_x$ (0.00 ≤ x ≤ 0.50) single crystals

P.K. Maheshwari[1,2], V. Raghavendra Reddy[3], Bhasker Gahtori[1] and V.P.S. Awana[1,2,*]

[1] CSIR-National Physical Laboratory, Dr. K. S. Krishnan Marg, New Delhi-110012, India
[2] Academy of Scientific and Innovative Research (AcSIR), Ghaziabad- 201002, India
[3] UGC-DAE Consortium for Scientific Research, Khandwa Road, Indore-425001, India

**Abstract**

Here, we report self flux single crystal growth of FeTe$_{1-x}$Se$_x$ (0.00≤ x ≤ 0.50) series via solid state reaction route; the resulted crystals as seen are shiny. X-Ray diffraction (XRD) performed on the surface of crystals elucidated the growth in (00*l*) plane, i.e. orientation in *c*-direction only. Scanning electron microscopy (SEM) images showed slab like morphology and EDX (Energy dispersive X-ray analyzer) confirmed that the crystals are closed to their designed compositions. Rietveld analysis of the XRD patterns of crushed crystal powders showed that the cell parameters decrease with Se content increase. Coupled magnetic/structural phase transition temperature, seen as a step in resistivity for the lower Se concentration i.e. 0.00 ≤ x ≤ 0.07, decreases from around 65K for x=0.0 to 50K for x=0.07 and it is not detected for higher x values. Superconductivity is observed by resistivity measurement for higher Se concentration i.e. 0.07 ≤ x ≤ 0.50, up to a maximum temperature of 14K at x=0.50. Thermally Activated Flux Flow (TAFF) analysis based on high field transport measurements in superconducting region done for x=0.20 crystal exhibited activated flux energy to be decreasing from 12meV (0.5Tesla) to 4.6meV (14Tesla). Raman spectroscopy at room temperature of synthesized samples exhibits all the allowed phonon modes with slight shift to higher frequency with Se content. Mossbauer spectra of FeTe$_{1-x}$Se$_x$ crystals series were recorded at 300 and 5K. At 5K, the average hyperfine field decreases systematically with Se content increase from 10.6 to 6.1Tesla for x=0.0 to x=0.20 samples. This indicates a possibility of co-existing magnetism and superconductivity in 0.07 ≤ x ≤ 0.20 crystals. For x=0.50 sample, no hyperfine field related to magnetic ordering is seen. Based on above results, detailed phase diagram of the FeTe$_{1-x}$Se$_x$ (0.00 ≤ x ≤ 0.50) compounds is defined in the present study.

*****Corresponding Author**
Dr. V. P. S. Awana:  E-mail: awana@nplindia.org
Ph. +91-11-45609357, Fax-+91-11-45609310
Homepage awanavps.webs.com

**Introduction**

Fe chalcogenide superconducting compounds [1, 2] possess the simplest PbO-type crystal structure among the Iron based superconducting family [1-6]. Because of their simplest crystal structure, having only Fe chalcogen layers, these are thought to be the ideal candidates for understanding the properties and mechanism of Fe based superconducting compounds [1-6]. The FeSe/As tetrahedral layers in Fe containing superconducting compounds are the key component, which play the role similar to that of $CuO_2$ layers in superconducting cuprates [7, 8]. As reported, doping of S/Se at Te site in PbO-type $FeTe_{1-x}Se_x$ suppresses the magnetic ordering of parent Anti-ferromagnetic (AFM) compound FeTe and induces superconductivity [9-13]. The ground state of Iron Chalcogenides i.e. FeTe compound orders anti-ferro-magnetically below 70K [9, 14-16]. At ambient conditions, maximum $T_c$ for Fe chalcogenides superconducting system is found to be 15K at x=0.50 [9, 16] and $T_c$ was increased to 37K by hydraulic pressure [17-18]. $T_c$ can be further increased up to 50K with favorable alkali metal (Na, Eu, Ca etc) intercalation [19-21].

Till now, the concept of superconductivity in Fe based compounds i.e. Fe chalcogenides and pnictides is not yet clear as $T_c$ for these compounds gets increased of up to 100K for FeSe mono layer [22, 23], which cannot be explained by electron-phonon coupling theory and hence are termed as unconventional superconductors [24,25]. Because the electron–phonon coupling in FeSe/Te system is found to be weak, the phenomenon behind its superconductivity is yet challenging. Raman spectroscopy on single crystals of appropriate quality may provide some important information related to the electron coupling in these types of compounds [26]. Further, because a single crystal has no grain boundary related problems and, hence, it is always desired by condensed matter physicists for physical property measurements and theoretical modelling.

Till date, FeSe based compounds has no universal accepted phase diagram because of its phase complexity, as, in this system several phases and different crystal structures are appeared simultaneously. According to the defined phase diagram of FeSe system, it is not easy to obtain the large-size single crystals of superconducting FeSe by the melt process technique [27]. On the other hand, the single crystal growth is relatively easy using simple melt process technique for Te substituted FeSe system [25, 28]. Here, we report the self flux growth of FeTe$_{1-x}$Se$_x$ (0.00 ≤ x ≤ 0.50) single crystals and their structural properties, superconducting property, Raman spectroscopy and Mossbauer spectra. When the temperature is decreased from 300K to 2K during resistivity measurements, the pure and some doped sample i.e. x=0 to 0.07 showed step like transition (coupled magnetic/structural) decreasing from 65K to 50K for x=0.00 to x=0.07, respectively. Other crystals show superconducting transition at low temperature and maximum $T_c$ (14K) is achieved for x=0.50 sample. Interestingly, the x=0.07 crystal showed both the magnetic/structural and superconducting transitions at temperature around 50 and 10K, respectively. In Raman spectra, all the phonon modes are identified in low frequency range with slight higher frequency shift with Se doping. Mossbauer spectra data showed that the higher Se doping suppress the magnetic ordering in FeTe$_{1-x}$Se$_x$ series and some crystals do have both magnetic and superconductivity co-existing. Worth mentioning is the fact that, beyond x=0.50, the crystals could not be grown and the easy and versatile simple melt process technique did not work.

**Experimental Details**

The FeTe$_{1-x}$Se$_x$ (0.00 ≤ x ≤ 0.50) single crystals series are grown in a common programmable automated furnace without any extra added flux and complicated heating algorithm. All the essential elements i.e., Fe, Se and Te powders of high purity (better than 4N) from Sigma Aldrich are taken in their stoichiometric ratio i.e., FeTe$_{1-x}$Se$_x$ (0.00 ≤ x ≤ 0.50) and grinded separately in high purity Ar gas filled glove box. The grinded powders are pelletized by hydrostatic pressure of 100kg/cm$^2$, and sealed separately in respective quartz tubes with the vacuum of better than 10$^{-3}$Torr. These sealed quartz tubes were kept in a programmable automated furnace and heated up to 1000$^0$C for 24 hours with an intermediate step of 450$^0$C at cooling rate of 2$^0$C/minute. Finally the furnace was cool down very slowly up to room temperature with the cooling rate of 10$^0$C/hour. The obtained crystals from furnace were very shiny. XRD of as obtained single crystals and their crushed powders was done using Rigaku X-ray diffractometer with CuK$_\alpha$ radiation of 1.541Å at room temperature. SEM image and EDX had been taken on ZEISS-EVO-10 electron microscope to understand the morphology and elemental composition of FeTe$_{1-x}$Se$_x$ single crystals. Electrical and magnetic measurements were performed on Quantum Design Physical Property Measurement System (PPMS-14Tesla) down to 2K and up to magnetic field of 14Tesla. Raman spectra were recorded for all the samples at ambient condition on Renishaw Raman spectrometer with excitation by laser beam of 514nm wavelength. [57]Fe Mossbauer measurements were performed on crushed powder samples having [57]Co radioactive source in transmission mode. Mossbauer spectra have been recorded using WissEl velocity drive, calibrated at room temperature with natural Iron absorber, and at low temperature measurements were performed using the Janis made superconducting magnet.

**Results and Discussion**

Single crystal XRD of FeTe$_{1-x}$Se$_x$ (0.00 ≤ x ≤ 0.50) series were performed at room temperature and obtained XRD patterns are shown in figure 1. From figure 1, it may be clearly seen that direction of crystal growth is in [00*l*] plane orientation i.e. in *c* direction, being as the highest peak intensities of XRD pattern are only occurring by (001), (002) and (003) planes, in *2θ* range of 10 to 55$^0$. This result impounds the single crystalline nature of the studied crystals. For generality, it can be mentioned that atom layer stock formation and XRD patterns of similar type were observed for other layered chalcogenide crystals [29-31]. Also, it can be concluded that *c* cell parameter decreases monotonically with increasing Se doping at Te site, as the position of (00*l*) plane is shifted to higher angle in *2θ* graph.

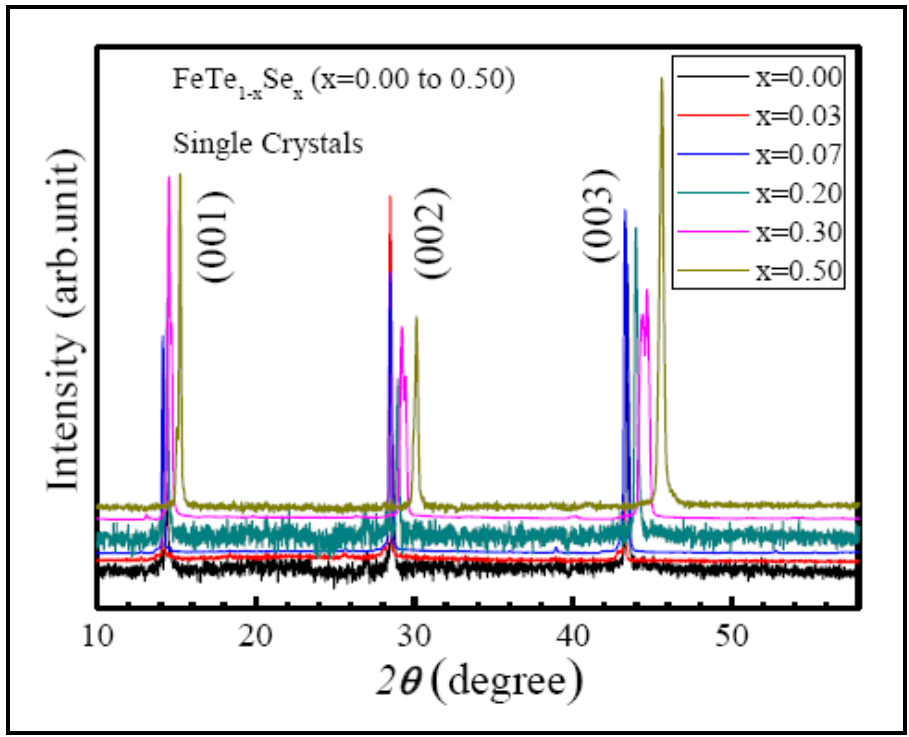

**Figure 1:** Single crystal XRD pattern for FeTe$_{1-x}$Se$_x$(x=0.00 to 0.50) series at ambient conditions.

**Table 1:** FeTe$_{1-x}$Se$_x$ (0.00 ≤ x ≤ 0.50) single crystals lattice parameters and coordinate positions using Rietveld refinement.

|  | **x= 0.00** | **x=0.01** | **x=0.03** | **x=0.05** | **x=0.10** | **x=0.30** | **x=0.50** |
|---|---|---|---|---|---|---|---|
| ***a=b*** (Å) | 3.8263(2) | 3.8267(4) | 3.8259(3) | 3.824(2) | 3.821(3) | 3.809(2) | 3.801(2) |
| ***c*** (Å) | 6.292(4) | 6.2858(3) | 6.2754(3) | 5.272(3) | 6.234(3) | 6.116(3) | 5.998(3) |
| **V(Å$^3$)** | 92.118(3) | 92.0491(3) | 91.8596(2) | 91.7467(2) | 91.04(2) | 88.7816(3) | 86.65(3) |
| **Fe** | (3/4,1/4,0) | (3/4,1/4,0) | (3/4,1/4,0) | (3/4,1/4,0) | (3/4,1/4,0) | (3/4,1/4,0) | (3/4,1/4,0) |
| **Se** | (1/4,1/4,0.286) | (1/4,1/4,0.289) | (1/4,1/4,0.288) | (1/4,1/4,0.289) | (1/4,1/4,0.283) | (1/4,1/4,0.284) | (1/4,1/4,0.286) |
| **Te** | (1/4,1/4,0.286) | (1/4,1/4,0.289) | (1/4,1/4,0.288) | (1/4,1/4,0.289) | (1/4,1/4,0.283) | (1/4,1/4,0.284) | (1/4,1/4,0.286) |

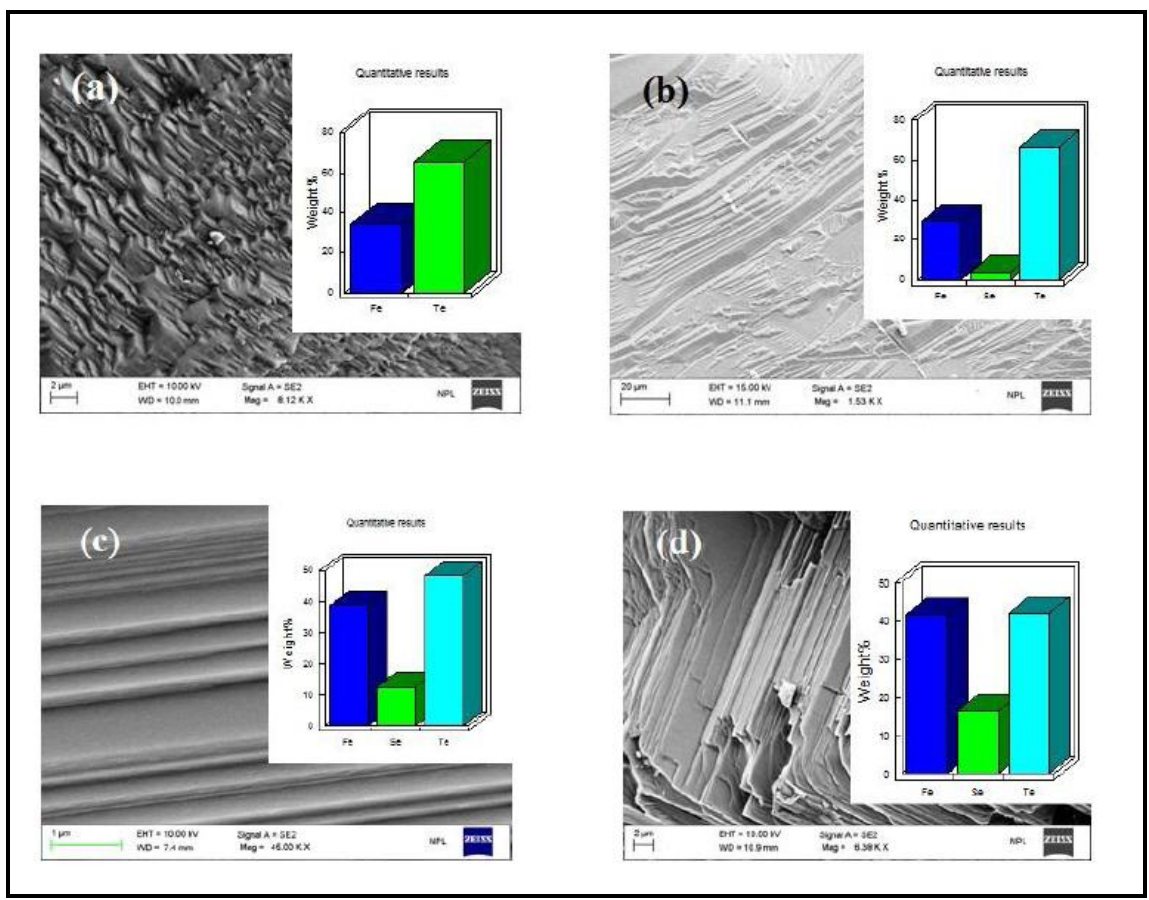

**Figure 2:** Selected SEM image of FeTe$_{1-x}$Se$_x$ crystals at room temperature (a) FeTe (b) FeTe$_{0.95}$Se$_{0.05}$ (c) FeTe$_{0.70}$Se$_{0.30}$ (d) FeTe$_{0.50}$Se$_{0.50}$. Insets are EDAX of all the samples respectively.

To evaluate the morphology of the synthesized single crystal, SEM observation was performed at room temperature. SEM images of FeTe$_{1-x}$Se$_x$ samples were taken at high magnification factor for understanding the morphology of the synthesized single crystals, and the results are shown in figure 2(a-d). The SEM image of pure FeTe is shown in Figure 2(a) and the quantitative analysis of the micro relief is presented in the inset. Although the SEM analysis of FeTe was reported elsewhere [32], It is shown for comparison with others crystals. Figure 2(b) and 2(c) show the SEM and EDX analysis of $FeTe_{0.95}Se_{0.05}$ and $FeTe_{0.70}Se_{0.30}$ crystals, respectively. The results show clearly the layered directionally parallel oriented morphology and elemental composition is found to be close to their stoichiometric ratio. Figure 2(d) depicts the earlier reported $FeTe_{0.50}Se_{0.50}$ sample [33] and it is essentially shown for comparison with the other crystals. From the SEM images and EDX, we can conclude that the morphology of all the samples is basically layered type and the elemental compositions are close to starting elemental ratios.

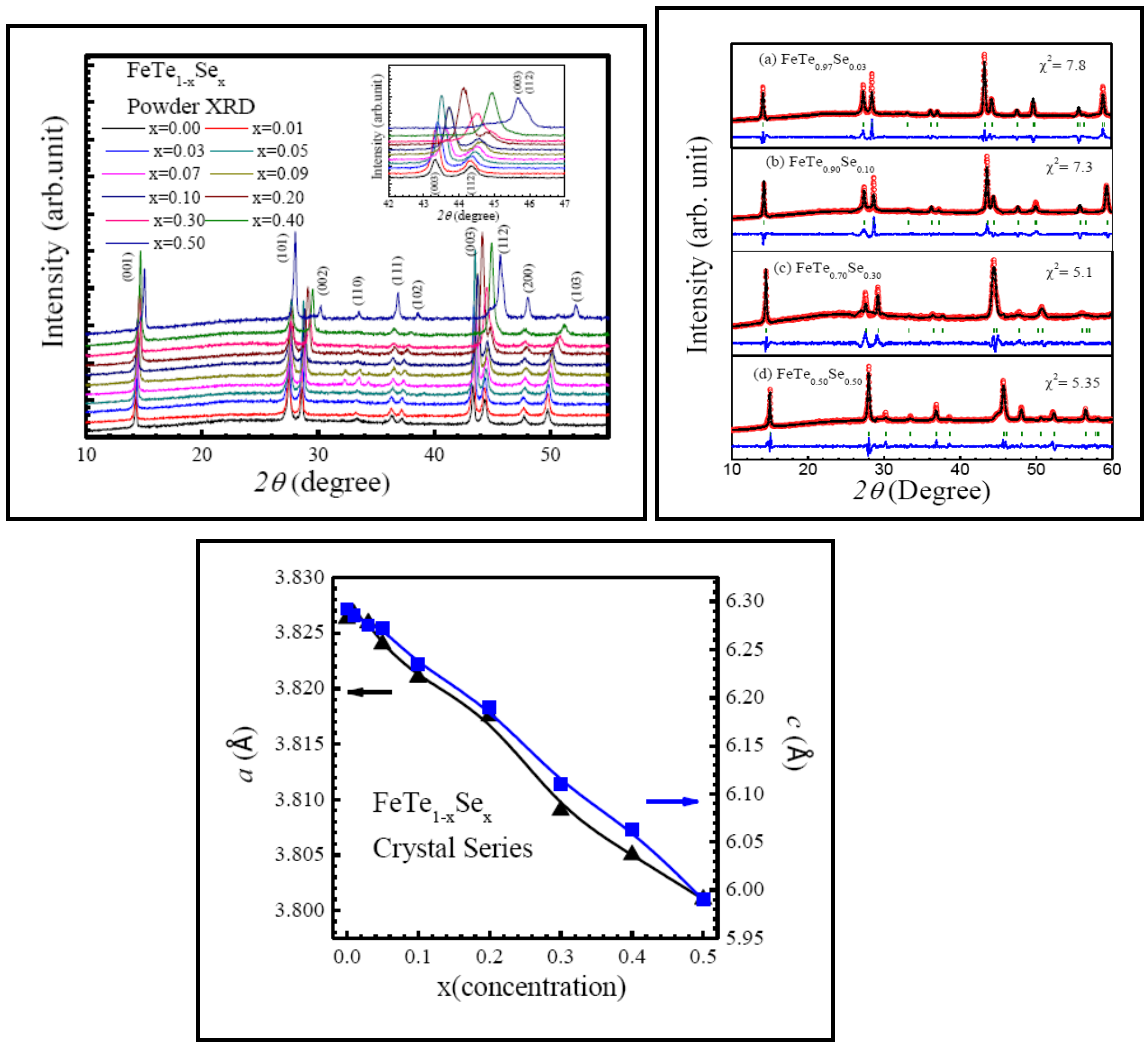

**Figure 3:** (a) Powder XRD of $FeTe_{1-x}Se_x$(x=0.00 to 0.50) series at room temperature. Inset is zoomed part of [003] plane. (b) Observed and Rietveld fitted graph of some of $FeTe_{1-x}Se_x$ series samples. (c) Plot between lattice parameters (*a* and *c*) with doping concentration(x) of $FeTe_{1-x}Se_x$ series.

For further analysis of the FeTe$_{1-x}$Se$_x$ (0.00 ≤ x ≤ 0.50) crystal series, concerning their cell parameters their lattice parameters, coordinate position etc., gently crushed powder XRD measurement was performed at room temperature, as shown in figure 3(a). All the samples exhibit the tetragonal type crystal structure, in space group P4/nmm without any impurity phase present in the samples. From the powder XRD result, it is clearly seen that (00$l$) peaks are shifted towards higher angle with Se concentration increase in FeTe$_{1-x}$Se$_x$ (0.00 ≤ x ≤ 0.50) series. This result confines the decrement in *c* lattice parameter with increasing Se partial occupancy at Te site. The inset of figure 3(a) clearly shows the shift of (003) peak towards higher angle.

The Rietveld refinement of powder XRD result was performed using fullprof software, and the observed and fitted plots of selected samples are shown in figure 3(b). The main refinement parameters are given in table 1. The results show the decrement in *a* and *c* parameters with increasing Se concentration in FeTe$_{1-x}$Se$_x$ series. As compared to *a* cell parameter, *c* cell parameter decreases sharply, which can be clearly seen from figure 3(c). In fig 3(c), it can be seen that both lattice parameters *a* and *c* decrease from 3.8263(2)Å to 3.801(2)Å and 6.292(4)Å to 5.998(3)Å respectively for x=0.00 to x=0.50. Respectively, fractional volume decreases with increasing Se concentration at Te site in FeTe$_{1-x}$Se$_x$ series.

For understanding the transport properties of the studied crystals, resistivity ($\rho$) verses temperature (*T*) measurements are performed from 300K to 2K for FeTe$_{1-x}$Se$_x$ (0.00 ≤ x ≤ 0.50) samples. The $\rho$-*T* dependencies are shown in figure 4(a). In FeTe$_{1-x}$Se$_x$ (0.00 ≤ x ≤ 0.50) series, for x=0.0 to 0.05, the known structural/magnetic ($T_N$) phase transition occurs in $\rho$-*T* measurement during cooling and warming cycles. For x=0, the transition occurs in resistivity measurement at ~65K, while for x=0.05 it occurs at ~53K. Further details about these crystals i.e. x=0.00 to 0.05 are shown elsewhere [32, 34]. There is no superconducting transition i.e., neither $T_c^{onset}$ nor $T_c^{offset}$ ($\rho$=0) occur for 0.0 ≤ x ≤ 0.05 samples. Interestingly, for x=0.07 both structural/magnetic and superconducting transition $T_c^{onset}$ occurs approximately at 50K and 10K respectively. The superconducting transition temperature $T_c$ increases from 10K for x=0.07 to 14K for x=0.50 with Se doping in FeTe$_{1-x}$Se$_x$ series.

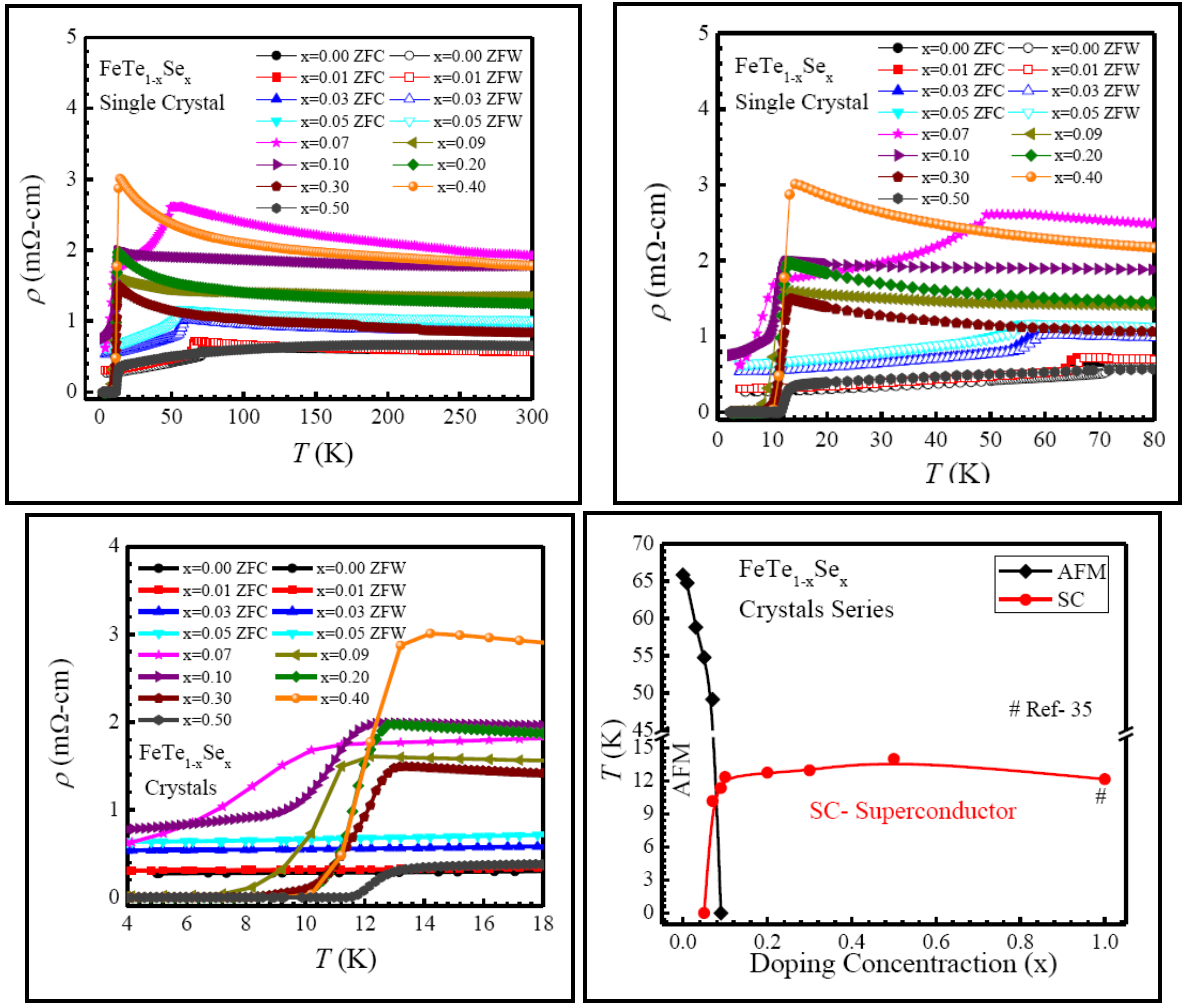

**Figure 4:** Electrical resistivity ($\rho$) versus temperature ($T$) plots for FeTe$_{1-x}$Se$_x$ (x=0.00 to 0.50) series (a) Temperature range from 300K to 2K. (b) Temperature range from 80K to 2K. (c) Temperature range from 18 to 4K. (d) Phase diagram of FeTe$_{1-x}$Se$_x$ (x=0.00 to 1.00) crystal series, which shows anti-ferromagnetism (AFM) and superconducting (SC) regions.

Figure 4(b) shows zoomed part of figure 4(a) in the range of 80 to 2K. In this figure, one can clearly see that for x=0.00, $T_N$ occurs nearly at around 70K in $\rho$-$T$ measurement with hysteresis of nearly 5K during cooling and warming cycle, being occurring due to presence of latent heat during transition. It is clear from figure 4(b) that $T_N$ decreases with increasing Se up to x=0.07 in FeTe$_{1-x}$Se$_x$ series and higher concentration lead towards superconducting transition at low temperature. From figure 4(b), we can conclude that up to x=0.05 neither $T_c^{onset}$ nor $T_c^{offset}$ ($\rho$=0) is seen down to 2K while for x=0.07 only $T_c^{onset}$ is occurred at nearly 10K. Also the hysteresis in $\rho$-$T$ measurement near $T_N$ is seen up to x=0.05 crystals during cooling/warming cycles, some of these results are reported earlier [32, 34]. Here, the aim is to fill up the gaps and present the complete phase diagram of FeTe$_{1-x}$Se$_x$ (x=0.0 to 0.50) single crystal series.

**Table 2:** FeTe$_{1-x}$Se$_x$(x=0.00 to 0.50) single crystals, phase transition temperature ($T_N$) and superconducting transition temperature ($T_c^{onset}$)

| x | $T_N$ (K) | $T_c^{onset}$ (K) |
|---|---|---|
| 0.00 | 65.84 | - |
| 0.01 | 64.77 | - |
| 0.03 | 58.81 | - |
| 0.05 | 54.78 | - |
| 0.07 | 49.17 | 10.18 |
| 0.10 | - | 11.38 |
| 0.20 | - | 12.37 |
| 0.30 | - | 12.76 |
| 0.50 | - | 12.97 |
| #1.00 | - | 12.17 |

*# Ref- 35*

Figure 4(c) shows the $\rho$-$T$ measurement from 18 to 4K, i.e. zoomed part of figure 4(a). From this figure, we can say that that maximum $T_c$ is found for x=0.50 sample i.e. FeSe$_{0.50}$Te$_{0.50}$ crystal. The $T_c^{onset}$ for x=0.50 is at nearly 14K and $T_c^{offset}$ ($\rho$=0) is at 12K. More detailed analysis of FeSe$_{0.50}$Te$_{0.50}$ single crystal was reported earlier [33]. The values of phase transition temperature ($T_N$) and superconducting transition temperature ($T_c^{onset}$) of FeTe$_{1-x}$Se$_x$ series are shown in table 2. From Table 2, we conclude that maximum $T_N$ is found nearly 65K for x=0.00 and maximum $T_c$ is found at 14K for x=0.50 crystal. The suggested temperature-doping concentration phase diagram of FeTe$_{1-x}$Se$_x$(x=0.00 to 1.0) series from our results is shown in figure 4(d). The detailed analysis of x=1.0 sample is reported earlier [35], and the related point is marked by # in fig 4(d) also. From fig 4(d), one can say that for x≤0.07, the FeSe/Te system shows the anti-ferromagnetism (AFM) ordering and, for x≥0.07 this system shows the superconducting (SC) state at low temperature. Interestingly, the x=0.07 shows both AFM and SC state at nearly 50K and 10K, respectively. So, from the suggested phase diagram, as Se concentration increases in FeTe1$_{-x}$Se$_x$ series, AFM ordering suppresses and superconductivity appears. Also, the crystals with Se content of x=0.07 show both AFM and SC states.

To further study superconducting response of studied FeTe$_{0.80}$Se$_{0.20}$ single crystal, $\rho$-$T$ measurement was performed under applied magnetic up to 14Tesla. The similar data for x =0.50 crystal were already reported earlier [33]. Figure 5(a) shows the $\rho$-$T$ behavior of FeTe$_{0.80}$Se$_{0.20}$ single crystal under applied magnetic field up to 14Tesla. From figure 5(a), we can judge the robustness of superconductivity against the magnetic field as the $T_c^{\text{offset}}(\rho=0)$ is only decreased from 10K to 5.5K for zero and 14Tesla magnetic fields respectively. The calculated $dT_c/dH$ for FeTe$_{0.80}$Se$_{0.20}$ sample comes out to be around 0.32K/Tesla. The low $dT_c/dH$ value in the studied crystal signifies the superconductivity robustness to magnetic field up to 14Tesla.

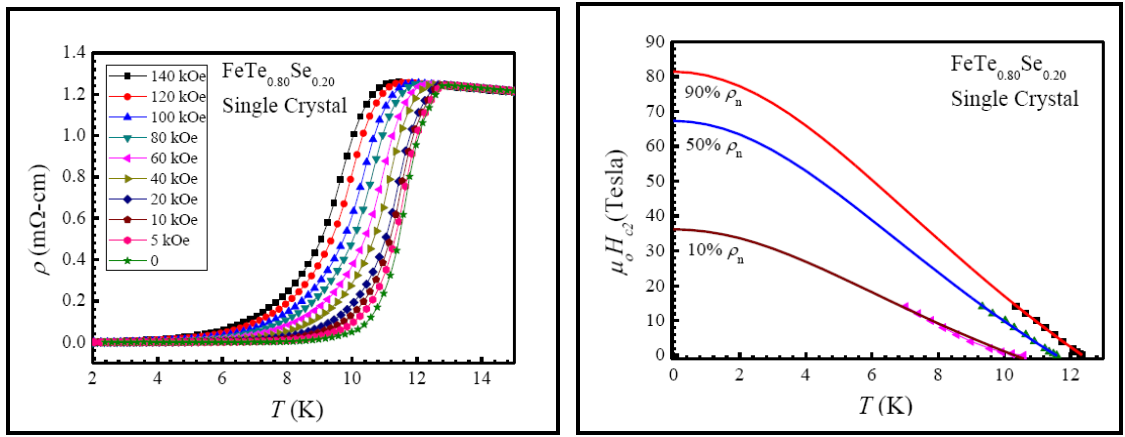

**Figure 5:** (a) $\rho(T, H)$ under various magnetic fields up to 14Tesla for FeTe$_{0.80}$Se$_{0.20}$ single crystal (b) Upper critical field ($H_{c2}$) is calculated from $\rho(T,H)$ data with 90%, 50% and 10% normalized resistivity ($\rho_n$) criteria of FeTe$_{0.80}$Se$_{0.20}$ single crystal.

For further analysis on $\rho(T, H)$ measurement for FeTe$_{0.80}$Se$_{0.20}$ crystal, the upper critical field $H_{c2}(0)$ is calculated using 10%, 50% and 90% normalized resistivity($\rho_n$) criterion, which is shown in figure 5(b). The $H_{c2}$ at absolute zero temperature is calculated by extrapolating the curve fitting the data using the Ginzburg Landau (GL) equation $H_{c2}(T) = H_{c2}(0)[(1 - t^2)/(1 + t^2)]$, here t = $T/T_c$ is defined as reduced temperature. The calculated values are found to nearly 80Tesla, 70Tesla and 35Tesla for FeTe$_{0.80}$Se$_{0.20}$ single crystal at 90%, 50% and 10% $\rho_n$ criterion respectively. High value of $H_{c2}$ for the superconductor leads towards the robustness against applied external magnetic field. The calculated value of $H_{c2}$ for FeTe$_{0.80}$Se$_{0.20}$ single crystal is very and it is far away from the Pauli Paramagnetic limit, i.e. 1.84 times of $T_c$ [36]. The coherence length [$\xi(0)$] can be calculated using the value of $H_{c2}(0)$ by the defined equation i.e. $H_{c2}(0)=\varphi_0/2\pi\xi(0)^2$, where $\varphi_0$ is define as flux quantum, whose value is 2.0678 X 10$^{-15}$ Tesla-m$^2$. At absolute zero temperature, $\xi(0)$ calculated by the equation is around 20.3Å for FeTe$_{0.80}$Se$_{0.20}$ single crystal.

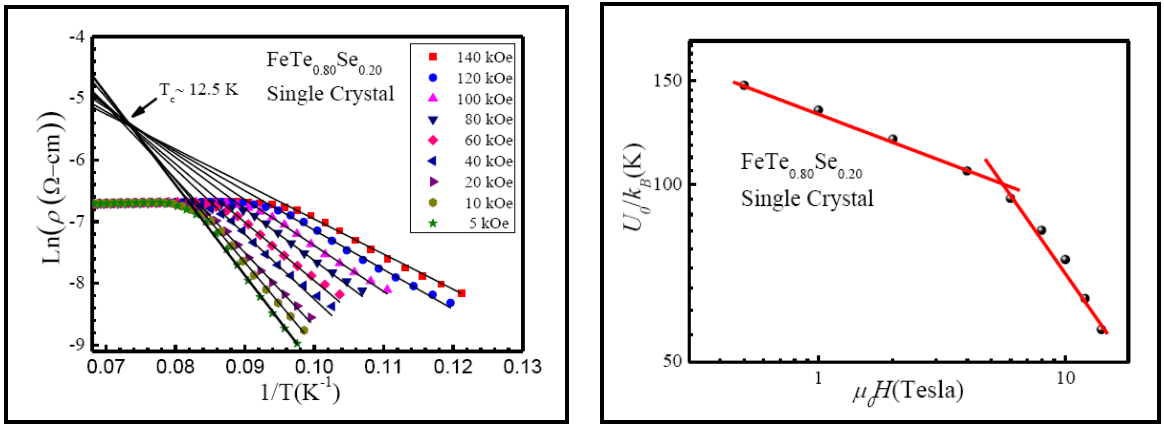

**Figure 6:** (a) ln$\rho$(T,H) vs 1/T for various magnetic fields up to 14Tesla for FeTe$_{0.80}$Se$_{0.20}$ single crystal corresponding linear fitted solid line of Arrhenius relation. (b) Thermally Activation energy $U_o(H)$ with solid lines fitted of $U_o(H) \sim H^\alpha$ up to 14Tesla magnetic field for FeTe$_{0.80}$Se$_{0.20}$ single crystal.

Then, the thermally activated flux flow (TAFF) analysis is implemented on FeTe$_{0.80}$Se$_{0.20}$ sample. The dependence of Ln$\rho$ verses 1/T under magnetic fields of up to 14Tesla is shown in figure 6(a). By the TAFF theory [37,38], the TAFF region can be described with the help of Arrhenius relation [39] Ln$\rho(T, H)$ = Ln$\rho_0(H)$ - $U_0(H)/k_BT$, where Ln$\rho_0(H)$ is temperature dependent constant, $U_0(H)$ is thermally activation energy and $k_B$ is the Boltzmann constant. From the Arrhenius equations we can conclude that Ln$\rho(T, H)$ verses 1/T graph would be linearly fitted in TAFF region, as shown in the figure 6(a). Figure 6(a) described that Ln$\rho(T, H)$ verses 1/T and linearly fitted plated in TAFF region up to 14Tesla magnetic field. In this figure, it can be seen that all the linear fitted extrapolation lines correspondent to magnetic field intercepted at same point, this point nearly coincides with superconducting transition temperature ($T_c$) of FeTe$_{0.80}$Se$_{0.20}$ sample, i.e. 13K. There is an occurrence of resistivity broadening pattern with applied magnetic field in FeTe$_{0.80}$Se$_{0.20}$ single crystal due to thermally assisted flux motion [40] and is very similar to our previous result for another superconducting FeSe$_{0.50}$Te$_{0.50}$ crystal [33].

The thermal activation energy was calculated for different magnetic fields for FeTe$_{0.80}$Se$_{0.20}$ single crystal, the Ln scale plot between $U_0(H)$ verses magnetic field is shown in figure 6(b). Thermally activation energy is field dependent energy which follows the power law relation with magnetic field, i.e. define as $U_0(H) = K \times H^\alpha$, where $U_0(H)$ is thermally activation energy, $K$ is constant and $\alpha$ is a field dependent constant. For the studied FeTe$_{0.80}$Se$_{0.20}$ single crystal the value of $\alpha$ is found 0.16 up to magnetic field of 4Tesla and 0.60 for higher magnetic fields from 6Tesla to 14Tesla range. This result indicates that single vortex pinning [38] is effective at low field for $FeTe_{0.80}Se_{0.20}$ single crystal. The thermal activation energy for $FeTe_{0.80}Se_{0.20}$ single crystal varies from 12 to 4.6meV with magnetic field from 0.5 to 14Tesla respectively. The calculated thermal activation energy for $FeTe_{0.80}Se_{0.20}$ crystal is quite low in comparison to $FeTe_{0.50}Se_{0.50}$ single crystal [33]. This shows that the superconductivity robustness for x=0.20 crystal is lower than that for x=0.50 [33].

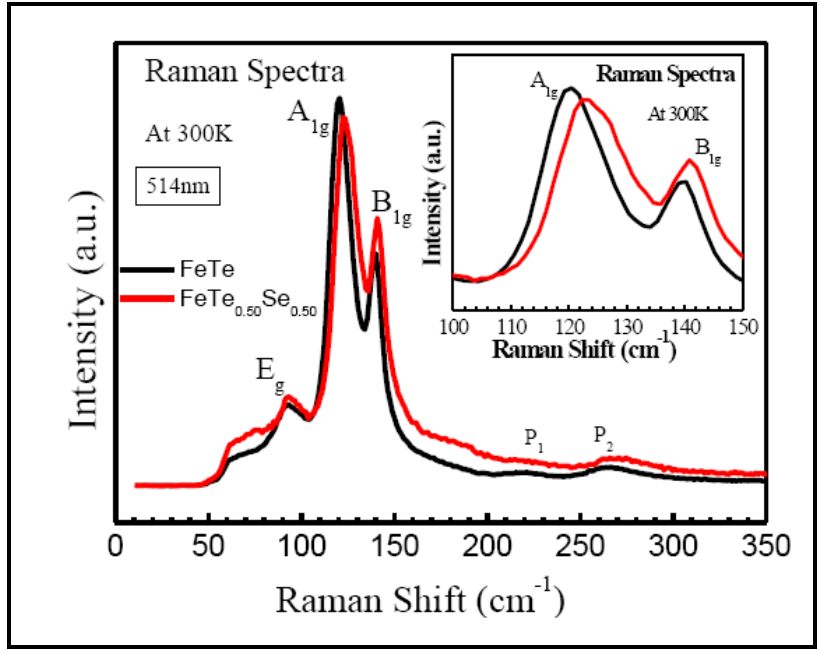

**Figure 7**: Raman spectra of $FeTe_{1-x}Se_x$ (x=0.00 and 0.50) series at room temperature. Inset is zoomed part of same.

In figure 7, the Raman spectra are shown as recorded for $FeTe_{1-x}Se_x$ (x=0.00 and 0.50) series at room temperature at the laser beam excitation of 514nm with laser power of 25mW. At this power Raman spectra shows the best signal/noise ratio. The Raman spectra were recorded in low wave number range of 50-350$cm^{-1}$. In the low frequency range for the Fe(Se/Te) system after taking symmetry concern, four Raman active modes ($E_g^1$, $A_{1g}$, $B_{1g}$ and $E_g^2$) are expected [41,42]. These modes $E_g^1$, $A_{1g}$, $B_{1g}$ and $E_g^2$ are attributed to Te/Se, Te/Se, Fe and Fe atom vibrations, respectively. Xia et al [41] explained the $A_{1g}$(Te) and $B_{1g}$(Fe) modes in low wave number range (100-200$cm^{-1}$) of FeTe crystal. In bulk $FeSe_{0.82}$ system, P. Kumar et al [42] observed two Raman modes attributed to $E_g$ (106$cm^{-1}$) and $A_{1g}$(160$cm^{-1}$) in the 100-200$cm^{-1}$ region. In frequency range of 100-200$cm^{-1}$, same peaks were observed by Lopes et al [26], as well as in $FeSe_{0.50}Te_{0.50}$ single crystal, at both 300 and 77K. In our Raman spectra recorded for FeTe and $FeTe_{0.50}Se_{0.50}$ crystals, we also found two Raman modes in low wave number region of 100-200$cm^{-1}$, that is very close to previous literature results for FeTe$_{0.50}$Se$_{0.50}$ [26]. In figure 7, the Raman spectra of FeTe and FeTe$_{0.50}$Se$_{0.50}$ crystals at room temperature, are shown in the wave number range of 50-200cm$^{-1}$. Three clear Raman modes are seen which are attributed to E$_g$(Te/Se), A$_{1g}$(Te/Se) and B$_{1g}$(Fe) at 92, 121 and 140cm$^{-1}$ respectively. Further, two weak Raman modes (P1 and P2) are observed in the range of 200-300cm$^{-1}$, which are similar to ones being observed by Lopes et al [26]. We also observed slight shifting in the positions of Raman active modes towards higher wave numbers with increasing Se concentration at Te sites in FeTe$_{1-x}$Se$_x$. In the inset of fig. 7, the zoomed part of wave number range (100-150cm$^{-1}$) is given to show the Raman peak shifting by Se doping. The Raman mode shifting may be due to lower atomic mass of Se, as compared to Te, and less strong bonding force of FeSe$_{0.50}$Te$_{0.50}$, as compared to FeTe.

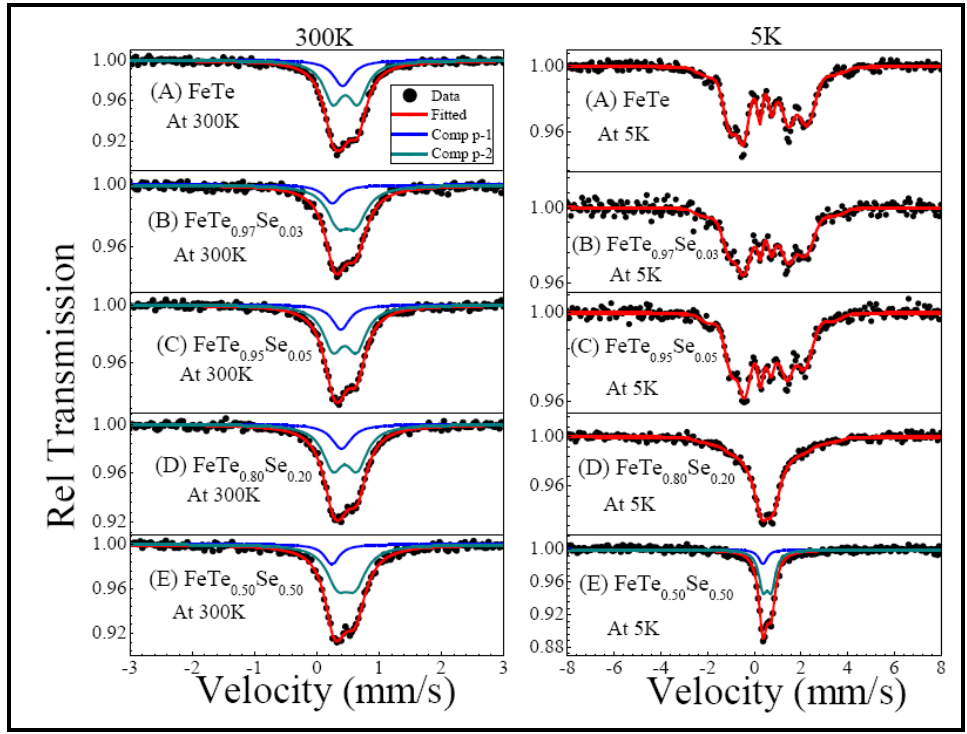

**Figure 8**: Mossbauer spectra of FeTe$_{1-x}$Se$_x$(x=0.00 to 0.50) crystal series, the left side frame is at 300K (RT) and right side is at 5K temperature.

The Mossbauer spectra of FeTe$_{1-x}$Se$_x$(x=0.00 to 0.50) at room temperature (RT, 300K) and lowest temperature of 5K are shown in left and right side frames of figure 8. For comparison, the Mossbauer spectra for x=0.00 sample from our previous study [43] are also provided. At the lowest measurement temperature (5K), the Mossbauer spectra display clear magnetic sextet till 0.05 and a broad doublet for x=0.20 indicating the presence of magnetic ordering in these crystals. Therefore, the 5K spectra of these crystals are analyzed with distribution of hyperfine fields and, for x=0.20 the data are analysed with doublet along with distribution of hyperfine fields.

The average hyperfine field is found to be 10.6(1), 10.2(1), 9.1(1) and 6.1(1) Tesla for x=0.0, 0.03, 0.05 and 0.20 samples, respectively. For higher doping (x=0.50), no sextet is observed at 5K, that can be attributed to clear absence of magnetic ordering in this crystal, as supported by $\rho(T)$ measurements. Clearly, successive doping of Se in FeTe results in suppression of magnetic nature, as revealed by Mossbauer spectra at 5K. The x=0.07 and 0.20 crystals both have magnetic and superconductivity co-existence.

**Conclusion**

The $FeTe_{1-x}Se_x$ ($0.00 \leq x \leq 0.50$) single crystals were successfully grown from self fluxes in a simple programmable furnace. The structural properties are characterized by XRD and SEM techniques. The cell parameters *a* and *c* decrease with Se concentration increase in $FeTe_{1-x}Se_x$ series. The magneto transport measurements down to 2K indicate that structural/magnetic phase transition occurs for $FeTe_{1-x}Se_x$ series up to x=0.07 and superconducting transition occurs for x≥0.07, although x=0.07 show both magnetic and superconducting transition at low temperatures. Highest $T_c$ of around 14K is found for x=0.50 sample. Raman and Mossbauer spectroscopy results are also reported for all the crystals and phase diagram of $FeTe_{1-x}Se_x$ single crystal series is completed by filling in the gaps with our previous studies [29-32].

**Acknowledgement**

Authors would like to thank their Director NPL India for his keen interest in the present work. This work is financially supported by the DAE-SRC outstanding investigator award scheme on search for new superconductors. P. K. Maheshwari thanks CSIR, India for research fellowship and AcSIR- Ghaziabad for Ph.D. registration. The authors would like to thanks Mrs. Shaveta Sharma for Raman studies.